\documentclass[preprint,1p,times]{elsarticle}
\usepackage{amssymb}
\usepackage{amsmath}
\usepackage{float}
\usepackage{comment}
\usepackage{titlesec}
\usepackage{xcolor}
\usepackage{multirow}

\titleclass{\subsubsubsection}{straight}[\subsection]

\newcounter{subsubsubsection}[subsubsection]
\renewcommand\thesubsubsubsection{\thesubsubsection.\arabic{subsubsubsection}}
\titleformat{\subsubsubsection}
  {\normalfont\normalsize\itshape}{\thesubsubsubsection}{1em}{}
\titlespacing*{\subsubsubsection}
{0pt}{3.25ex plus 1ex minus .2ex}{1.5ex plus .2ex}

\makeatletter
\renewcommand\paragraph{\@startsection{paragraph}{5}{\z@}%
  {3.25ex \@plus1ex \@minus.2ex}%
  {-1em}%
  {\normalfont\normalsize\itshape}}
\renewcommand\subparagraph{\@startsection{subparagraph}{6}{\parindent}%
  {3.25ex \@plus1ex \@minus .2ex}%
  {-1em}%
  {\normalfont\normalsize\itshape}}
\def\toclevel@subsubsubsection{4}
\def\toclevel@paragraph{5}
\def\toclevel@paragraph{6}
\def\l@subsubsubsection{\@dottedtocline{4}{7em}{4em}}
\def\l@paragraph{\@dottedtocline{5}{10em}{5em}}
\def\l@subparagraph{\@dottedtocline{6}{14em}{6em}}
\makeatother

\journal{TBD}
\begin{document}

\newcommand {\inred}[1] {\textcolor{red}{#1}}
\newcommand\XOR{\mathbin{\oplus}}
\newcommand\OR{\mid}
\newcommand\AND{\&}
\newcommand\myceil[1]{\left\lceil{#1}\right\rceil}
% SFMT notation: triple shifts operate on the full 128-bit element;
% plain \ll / \gg operate independently on each 32-bit sub-word.
\newcommand\SHL{\lll}
\newcommand\SHR{\ggg}

\begin{frontmatter}
\title{"VMT19937: A SIMD-Friendly Pseudo Random Number Generator based on Mersenne Twister 19937"}

\author{\renewcommand*{\thefootnote}{\fnsymbol{footnote}}
	Fabio Cannizzo\footnote{DISCLAIMER: Although Fabio Cannizzo is employed by Standard Chartered at the time this paper is written, this paper has been produced by Fabio Cannizzo in a personal capacity and Standard Chartered is not associated or responsible for its content.}}

\begin{abstract}
The Mersenne Twister (MT19937) is a widely used pseudo-random number generator, but its linear recurrence is not natively amenable to the SIMD parallelization available on modern hardware. This paper introduces a novel inter-state vectorization strategy that achieves near-perfect SIMD utilization by interleaving multiple MT instances de-phased via jump-ahead transformations. We apply this approach to the MT19937-32, MT19937-64, and SFMT19937 variations of the MT19937 algorithm, yielding the \textbf{V-family} of generators. This strategy effectively eliminates the alignment and dependency obstacles inherent in traditional intra-state vectorization. Benchmarks across multiple architectures, including Intel AVX-512 and ARM SVE, demonstrate that throughput scales linearly with SIMD register width. Our optimized implementations deliver speedups of an order of magnitude over reference scalar versions and outperform proprietary well-known libraries, while preserving the statistical properties and period of the underlying generators.
\end{abstract}

\begin{keyword}
PRNG \sep MT19937 \sep SIMD \sep vectorization
\end{keyword}

\end{frontmatter}

%----------------------------------------------------------------------------------------------------------%
\section{Introduction}
\label{sec:introduction}
Many simulation applications necessitate the generation of long sequences of pseudo-random numbers. These sequences appear statistically random despite being produced by deterministic and repeatable algorithms, effectively approximating some statistical properties of true randomness.

Various algorithms for pseudo-random number generators (PRNGs) are available, each differing in complexity and approximating distinct properties of true randomness. In specific application domains, the simplest and fastest PRNG that satisfies the required randomness properties is typically adopted. For example, Monte Carlo simulations demand long sequences of random numbers adhering to the central limit theorem hypothesis, requiring independence and identical distribution among the numbers. Linear recurrences modulo 2 serve as widely used building blocks for constructing PRNGs meeting such requirements. They involve a long binary state vector that evolves iteratively through linear recurrences.

One widely accepted pseudo-random generator from this family is the Mersenne twister 19937 (MT19937), proposed by Matsumoto and Nishimura in 1998 \cite{mt19937}. It has found implementation in several software libraries and numerical packages such as Matlab, Octave, Numerical Algorithms Group (NAG), Intel Math Kernel Library (MKL) and the C++ Standard Template Library (STL). Despite the proposal in the literature of more modern generators, such as XORSHIFT \cite{xorshift} and WELL \cite{well}, at present MT19937 remains de facto the most utilized PRNG (at least in the financial domain).

In addition to the good quality of its statistical properties, the success of MT19937 can be attributed to the simplicity of its algorithm. The linear recurrence responsible for evolving the binary state vector can be efficiently implemented with a small set of elementary bit manipulations, resulting in fast execution. However, this recurrence is not amenable to exploiting SIMD operations available on modern hardware for further speed enhancement.

To address this limitation, Saito and Matsumoto proposed SFMT19937 \cite{sfmt19937}. Rather than being a vectorized version of the original MT19937, SFMT is a distinct algorithm where the internal state and transition functions have been specifically re-calibrated for 128-bit words. While it naturally exploits 128-bit registers (e.g., SSE, NEON), it remains a strictly sequential algorithm that is difficult to vectorize further. Its design is hard-coded for a 128-bit word width, making it not generically adaptable to modern architectures with wider register lengths, such as 256-bit (Intel AVX) or 512-bit (Intel AVX-512).

This paper introduces the \textbf{V-family}: a new set of SIMD-friendly random number generators constructed by merging multiple instances of an existing base generator into a single, synchronized state. By de-phasing these instances via jump-ahead transformations and polling them in round-robin fashion as part of the same algorithm, we preserve the statistical properties and period of the underlying base generator while enabling full hardware utilization.
Because these instances remain synchronized, their state vectors can be evolved simultaneously via SIMD operations, achieving perfect inter-state vectorization.

As discussed in \cite{multistream}, the idea of splitting a PRNG sequence into multiple long sub-sequences has already been used to generate independent streams for parallel computing; the novelty of this work lies in exploiting this idea to achieve SIMD vectorization within a single synchronized algorithm.

The approach is applied to three base generators --- MT19937, MT19937-64, and SFMT19937 --- yielding the \textbf{V-family}: V-MT19937, V-MT19937-64, and V-SFMT19937. Benchmark comparisons also include a set of optimized intra-state SIMD implementations developed as part of this work (the X-family: X-MT19937, X-MT19937-64, X-SFMT19937) and Intel MKL, both of which serve as performance baselines.

\section{Mersenne Twister family}
\label{sec:mt_family}
\subsection{Common framework}
The MT19937 is a pseudo random number generator proposed by Matsumoto and Nishimura \cite{mt19937}.
Given an initial state $X = (x_0, x_1, \dots, x_{n-1})$ consisting of $n$ words of $w$ bits each, the generator produces a sequence $\boldsymbol{Z}$ of pseudo-random words $z_k$ with uniform distribution in $[0, 2^w-1]$. These words are generated sequentially for $k=0, \dots, P-1$, where $P$ is the period after which the sequence repeats, using the following algorithm:
\begin{equation}
\label{eq:formula}
\begin{aligned}
    x_{k+n} &= f(x_{k}, x_{k+1},\dots, x_{k+n-1}) \\
    z_k &= g(x_{k+n}) \\
\end{aligned}
\quad k=0\dots P-1
\end{equation}
To describe the recurrence $f(\cdot)$ and the transformation $g(\cdot)$ used in \eqref{eq:formula}, we first introduce some notation. Let $u$ and $v$ be words of $w$ bits, $\sigma$ be an integer $0\le \sigma \le w$, and $A$ be a binary matrix of size $w \times w$:
\begin{itemize}
    \item $(u \OR v)$ is the bitwise OR of $u$ and $v$
    \item $(u \XOR v)$ is the bitwise XOR of $u$ and $v$
    \item $(u \AND v)$ is the bitwise AND of $u$ and $v$
    \item $(u \gg \sigma)$ is the bitwise SHIFT RIGHT of $u$ by $\sigma$ positions
    \item $(u \ll \sigma)$ is the bitwise SHIFT LEFT of $u$ by $\sigma$ positions
    \item $(u~\%~v)$ is the remainder of the integer division of $u$ and $v$
    \item $u \star A$ is the binary matrix-matrix product in modulo-2, where $u$ is interpreted as a row binary matrix
\end{itemize}
Assuming that $\AND$ has higher priority than $\OR$ and $\star$ has higher priority than $\XOR$, the recurrence $f(\cdot)$ is given by:
\begin{equation}
\label{eq:step}
    x_{k+n} = f(x_{k},x_{k+1}, \dots, x_{k+n-1}) = x_{k+m} \XOR \underbrace{\left(x_k \AND h \OR x_{k+1} \AND l\right)}_{\omega}\, \star \,A
\end{equation}
where the recurrence parameters are defined as follows:
\begin{itemize}
	\item $m$ is an index chosen in the range $0 \le m < n$
    \item $h$ and $l$ are $w$-bits constant words defined as $h=\sum_{i=r}^{w-1}2^i$ and $l=\sum_{i=0}^{r-1}2^i$, for some chosen integer $0 \le r \le w-1$
    \item $A$ is a constant binary matrix of size $w \times w$
\end{itemize}

The matrix $A$ has a particular structure chosen so that the vector matrix multiplication $(\omega\star A)$ can be carried out quickly with just a few simple bit manipulations. Let $(a_0, a_1, \dots, a_{w-1})$ the individual bits of some constant word $a$, where $a_0$ is the least significant bit, and $I_{w-1}$ a binary identity matrix of size $(w-1)$
\begin{equation}
\label{eq:matmult}
A = \left[ \begin{matrix} 0 & I_{w - 1} \\ a_{w-1} & (a_{w - 2}, \ldots , a_0) \end{matrix} \right] \quad \implies \quad \omega\star A = \begin{cases}\omega \gg 1 & \text{if $\omega$ is even}\\(\omega \gg 1) \oplus a & \text{if $\omega$ is odd}\end{cases}
\end{equation}

Note that although the stored state vector has size $nw$ bits, the lower $r$ bits of $x_k$ are not used in recurrence \eqref{eq:step} and after the recurrence is complete they are discarded, so the effective dimension of the state vector is only $(nw-r)$ bits. \\ % This determines the period of the generator $P=2^{nw-r}-1$.

The transformation $g(\cdot)$ in \eqref{eq:formula}, called \textit{tempering}, is also chosen to be quickly computable. It is defined as the following sequential manipulation steps:
\begin{equation}
\label{eq:tempering}
\begin{aligned}
y &= x_{k+n} \oplus ((x\gg \alpha)\!\And\! d)\\
y &= y \oplus ((y\ll \beta)\!\And \!b)\\
y &= y \oplus ((y\ll \gamma)\!\And \!c)\\
z_k &= y \oplus (y\gg \delta)
\end{aligned}
\end{equation}
Here, $d$, $b$, and $c$ are $w$-bit constant words, while $\alpha$, $\beta$, $\gamma$, and $\delta$ are non-negative integer constants. The variable $y$ is a temporary $w$-bit auxiliary word used throughout the transformation.

The first two variants below (MT19937-32 and MT19937-64) share this common structure and differ only in their choice of parameters and word width $w$. The third variant (SFMT19937) uses a different recurrence that does not follow this template.

\subsection{MT19937 (32-bit)}
\label{sec:mt32}
The original MT19937 \cite{mt19937} uses 32-bit words.
The chosen parameters are:
\begin{equation}
\label{eq:params}
\begin{aligned}
(w, n, m, r) &= (32, 624, 397, 31)\\
a &= \text{0x9908B0DF}\\
(\alpha, d) &= (11, \text{0xFFFFFFFF})\\
(\beta, b) &= (7, \text{0x9D2C5680})\\
(\gamma, c) &= (15, \text{0xEFC60000})\\
\delta &= 18\\
\end{aligned}
\end{equation}
where $a$, $d$, $b$ and $c$ are given in hexadecimal format using \textit{C}-language notation.
Note that $d = \text{0xFFFFFFFF}$, so in the first tempering step the application of the AND mask can be skipped.
These parameters yield a period $P=2^{nw-r}\!-\!1=2^{19937}\!-\!1$ and good $K$-distribution properties \cite{mt19937}.

\subsection{MT19937-64 (64-bit)}
\label{sec:mt64}
The MT19937-64 \cite{mt19937_64} is the 64-bit variant proposed by Nishimura. It uses the same recurrence \eqref{eq:step} and tempering \eqref{eq:tempering}, but with parameters:
\begin{equation}
	\label{eq:params64}
	\begin{aligned}
		(w, n, m, r) &= (64, 312, 156, 31)\\
		a &= \text{0xB5026F5AA96619E9}\\
		(\alpha, d) &= (29, \text{0x5555555555555555})\\
		(\beta,  b) &= (17, \text{0x71D67FFFEDA60000})\\
		(\gamma, c) &= (37, \text{0xFFF7EEE000000000})\\
		\delta &= 43
	\end{aligned}
\end{equation}
While the 64-bit variant shares the same upper/lower bit split ($r=31$) as the 32-bit version, its wider $w=64$ word size results in a different masking configuration. Specifically, $\omega$ in \eqref{eq:step} draws 33 bits from $x_k$ (the upper mask $h$) and 31 bits from $x_{k+1}$ (the lower mask $l$), rather than the 1-bit/31-bit split of the original version.
Consequently, although the 64-bit variant uses half as many state words ($n=312$), the total state size remains $nw - r = 19937$ bits, preserving the period $P = 2^{19937}-1$. Each recurrence advances the state by one 64-bit word and extracts a single random value. Notably, because the mask $d$ is not trivial (i.e., $d \neq 0xF\!\dots F$), all four tempering steps require non-trivial masking, in contrast to the 32-bit variant.

\subsection{SFMT19937 (128-bit SIMD-oriented variant)}
\label{sec:sfmt}
The SFMT19937 (SIMD-oriented Fast Mersenne Twister) \cite{sfmt19937} was proposed by Saito and Matsumoto to exploit 128-bit SIMD arithmetic natively. Unlike the variants above it does \emph{not} follow the recurrence template \eqref{eq:step}. The state consists of $N=156$ elements $\mathbf{w}_k$, each a 128-bit integer viewed as a vector of four 32-bit sub-words $\mathbf{w}_k = (w_k^{(0)}, w_k^{(1)}, w_k^{(2)}, w_k^{(3)})$, for a total of $N \times 128 = 19968$ bits, of which only 19937 are effective.

We write $\mathbf{w} \SHL p$ for a left shift of the \emph{full} 128-bit element by $p$ bits, $\mathbf{w} \SHR p$ for the corresponding right shift, $\mathbf{w} \ll p$ for an \emph{independent} left shift of each 32-bit sub-word by $p$ bits, and $\mathbf{w} \gg p$ for the independent right shift. The recurrence is:
\begin{equation}
\label{eq:sfmt_step}
\mathbf{w}_{k+N} = \mathbf{w}_k
    \oplus (\mathbf{w}_k \SHL 8)
    \oplus \bigl((\mathbf{w}_{k+m} \gg 11)\, \AND\, \mathbf{K}\bigr)
    \oplus (\mathbf{w}_{k+N-2} \SHR 8)
    \oplus (\mathbf{w}_{k+N-1} \ll 18)
\end{equation}
where $\mathbf{K}=(K_1, K_2, K_3, K_4)$ is a 128-bit constant mask, and the parameters are:
\begin{equation}
\label{eq:sfmt_params}
\begin{aligned}
(w, N, m) &= (128,\; 156,\; 122)\\
(K_1,\, K_2,\, K_3,\, K_4) &= (\text{0xDFFFFFEF},\; \text{0xDDFECB7F},\; \text{0xBFFAFFFF},\; \text{0xBFFFFFF6})
\end{aligned}
\end{equation}

SFMT19937 shares the same huge period of $2^{19937}-1$ as the original Mersenne Twister. It does not need a separate tempering step, because its complex multi-lane mixing ensures the output is well-distributed and statistically sound right from the start.

\subsection{Implementation}
For each new random word generated the state vector is transformed by the recursion $$x_{k+n} = f(x_k, \dots, x_{k+n-1})$$ removing the least significant word $x_{k}$ on its left side and adding a new word $x_{k+n}$ on its right side.
Starting with an initial state vector $X_0=(x_0, x_1, \dots, x_{n-1})$, the state vector evolves as
$$
\begin{matrix}
X_0 &=& x_0 & x_1 & \dots & x_{n-1} &\\
X_1 &=& & x_1 & x_2 & \dots & x_{n} & \\
X_2 &=& & & x_2 & \dots & \dots & x_{n+1} &\\
\cdots \\
\end{matrix}
$$
This can be effectively implemented as a circular buffer of size $n$, which avoids shifting the position of each word inside the vector at each iteration.
\begin{equation}
\label{eq:staterec}
\begin{array}{ccccccccl}
X_0 &=& x_0 & x_1 & x_2 & \dots & x_{n-1} \\
X_1 &=& x_{n} & x_1 & x_2 & \dots & x_{n-1} & \quad&\text{(the new element $x_n$ overwrites $x_0$)} \\
X_2 &=& x_{n} & x_{n+1} & x_2 & \dots & x_{n-1} & \quad&\text{(the new element $x_{n+1}$ overwrites $x_1$)}\\
\cdots \\
\end{array}
\end{equation}
The recurrence can be rewritten in terms of the indices of the circular buffer by taking the modulo-$n$ of all indices.
\begin{equation}
\label{eq:stepmod}
x_{k\%n} = x_{(k+m)\% n} \XOR \dots
\end{equation}
Note that after $n$ steps, i.e. repeating \eqref{eq:stepmod} $n$ times, all elements of the state vector are replaced by new ones. To remove the modulo-$n$ operator from the buffer indices, the loop which generates an entirely new state vector can be broken in 3 sub-loops, depending on when the 3 indices $k$, $k+1$ and $k+m$ become equal to $n$.
\begin{equation}
\label{eq:stepnomod}
	x_k = \left\{
		\begin{aligned}
			x_{k+m}   & ~~\XOR & \dots &\quad &k = 0, \dots, n-m-1\\
			x_{k+m-n} & ~~\XOR & \dots &\quad &k = n-m, \dots, n-2 \\
			x_{k+m-n} & ~~\XOR & \dots &\quad &k = n-1
		\end{aligned}
    \right.
\end{equation}
After the state vector has been advanced by $n$ steps, $n$ new pseudo random words can be produced by applying transformation \eqref{eq:tempering} (or the SFMT-specific mixing) to the new state elements.
The three-sub-loop decomposition \eqref{eq:stepnomod} applies to all MT variants; the concrete values of $(n,m)$ are: $(624, 397)$ for MT19937-32, $(312, 156)$ for MT19937-64, and $(156, 122)$ for SFMT19937.

\subsection{Vectorization obstacles for MT19937 and MT19937-64}
\label{sec:vectorization}
Two fundamental obstacles arise when attempting to vectorize the state-vector update \eqref{eq:stepnomod}, and both affect all three generator variants.

The first obstacle is that the iteration count of the first (longest) sub-loop is generally not a multiple of the SIMD vectorization width $L/w$. For each variant the first sub-loop counts are: MT19937-32: $624-397=227$ iterations; MT19937-64: $312-156=156$ iterations; SFMT19937: $156-122=34$ iterations. Vectorization packs operations from consecutive loop iterations so that $L/w$ iterations execute simultaneously using SIMD instructions. As a concrete example, consider MT19937-32 with 128-bit SSE2 registers ($L=128$, $w=32$), which pack 4 operations at once. The first sub-loop could be implemented in vectorial format as:
\begin{align}
\label{eq:stepsse2}
\begin{pmatrix}x_{k} \\ x_{k+1} \\ x_{k+2} \\ x_{k+3} \end{pmatrix}
&= \begin{pmatrix}x_{k+m} \\ x_{k+1+m} \\ x_{k+2+m} \\ x_{k+3+m} \end{pmatrix} ~\XOR~ \left(\begin{pmatrix}x_{k} \\ x_{k+1} \\ x_{k+2} \\ x_{k+3} \end{pmatrix}~\AND~h ~\OR~ \begin{pmatrix}x_{k+1} \\ x_{k+2} \\ x_{k+3} \\ x_{k+4} \end{pmatrix}~\AND~l\right) \star A \quad && k=0, 4, \dots, 220 \\
\label{eq:stepssescalar}
x_k &= x_{(k+m)} ~\XOR~ \left(x_{k}~\AND~h ~\OR~ x_{k+1}~\AND~l\right) \star A \quad &&k = 224, 225, 226
\end{align}
Since 227 is not a multiple of 4, the best achievable is 56 packs of 4 iterations as in \eqref{eq:stepsse2}, followed by 3 scalar iterations as in \eqref{eq:stepssescalar}. Analogous remainders arise for any register width and for all three generator variants.

The second obstacle is that consecutive groups of state words required by adjacent iterations of \eqref{eq:stepnomod} are not memory-aligned. Continuing the MT19937-32 example: if $(x_k, x_{k+1}, x_{k+2}, x_{k+3})$ starts at a 16-byte-aligned address, then $(x_{k+1}, x_{k+2}, x_{k+3}, x_{k+4})$ starts 4 bytes later and is therefore misaligned. Similarly, $(x_{k+m}, x_{k+1+m}, x_{k+2+m}, x_{k+3+m})$ starts at an offset of $m \times 4 = 397 \times 4 = 1588$ bytes from the beginning of the state vector, which is not a multiple of 16. Misaligned SIMD loads which cross cache lines incur a performance penalty on modern CPUs (see for example \cite{intel} sections 6.3, 15.6 and 18.23 for X86 machines, or \cite{arm} for ARM-specific details).

Both obstacles are inherent to the sequential structure of the state recurrence and cannot be fully eliminated without changing the algorithm's mathematical structure.

For SFMT19937 vectorization is even harder. While its use of 128-bit words makes it a natural fit for 128-bit SIMD instructions, it does not resolve the broader vectorization obstacles for wider hardware. Its mathematical structure remains fundamentally capped at a 128-bit intra-state width; the unbreakable recurrence chain between consecutive words (i.e. the dependence of $x_n$ on $x_{n-1}$) prevents any throughput gain on 256-bit or 512-bit registers without the inter-state strategy introduced in the next section.

\section{Inter-State Vectorization: the V-Family}
\label{sec:simdgen}
In this section, we introduce a family of novel random number generators --- the \textbf{V-family} --- each built from multiple instances of a base MT generator combined in a way that is specifically designed to fully exploit SIMD hardware. These generators preserve the same statistical properties and period as the underlying base generator, and their theoretical throughput scales proportionally to the SIMD register width. The construction applies uniformly to all three base generators: MT19937-32, MT19937-64, and SFMT19937.

Given a certain initial state $X_0$, a base MT PRNG produces a sequence $\boldsymbol{Z}$ of PRNs:
\begin{equation}
\label{eq:mainseq}
   \boldsymbol{Z} = z_0, z_1, \dots, z_{P-1}
\end{equation}
The PRNs $z_k$ are independent and identically distributed, or at least they emulate such statistical properties. They have uniform distribution over words of width $w$, and the sequence has a period of $P=2^{19937}-1$.

Let's consider $M$ sub-sequences $\boldsymbol{Z_t}$ ($t=0\dots M-1$) of equal size $J$ obtained by partitioning sequence $\boldsymbol{Z}$ in groups of equal length $J=P/M$
\begin{equation}
\label{eq:subseq}
   \underbrace{z_0, z_1, \dots, z_{J-1}}_{\boldsymbol{Z_0}},\, \underbrace{z_J, z_{J+1}, \dots, z_{2J-1}}_{\boldsymbol{Z_1}}, \dots,\, \underbrace{z_{(M-1)J}, z_{(M-1)J+1}, \dots, z_{MJ-1}}_{\boldsymbol{Z_{M-1}}}
\end{equation}
Let's construct a new sequence $\boldsymbol{S}$ which interleaves the sequences $\boldsymbol{Z_t}$
\begin{equation}
\label{eq:combseq}
   \boldsymbol{S}=\underbrace{z_0, z_J, z_{2J}, \dots, z_{(M-1)J}}_{\text{1-st number from each sub-sequence}}, \,\underbrace{z_1,\, z_{J+1}, \, z_{2J+1},\, \dots,\, z_{(M-1)J+1}}_{\text{2-nd number from each sub-sequence}}\,, \dots\,, \underbrace{z_{J-1}, z_{2J-1}, z_{3J-1}, \dots, z_{MJ-1}}_{\text{$J$-th number from each sub-sequence}}
\end{equation}
In other words, sequence $\boldsymbol{S}$ is obtained by polling each sub-sequence $Z_t$ in round-robin fashion. This technique is known as the leap-frog method \cite{multistream}. Because the numbers in the original sequence $\boldsymbol{Z}$ are independent and identically distributed, the numbers in the new sequence $\boldsymbol{S}$, obtained by interleaving the sub-sequences $\boldsymbol{Z_t}$, are also independent and identically distributed. The combined sequence $\boldsymbol{S}$ preserves the same overall period and statistical properties as the original MT sequence.

Let $L$ be the SIMD register width in bits available on a given CPU (128, 256, or 512 depending on whether the ISA supports SSE/NEON, AVX/SVE, or AVX-512), and let $w$ be the state word size of the base generator ($w=32$ for MT19937-32, $w=64$ for MT19937-64, $w=128$ for SFMT19937). We choose $M=L/w$ as the \emph{vectorization coefficient}, i.e.\ the number of independent generator instances that can be evolved simultaneously in a single set of SIMD operations.
We divide the total period $P=2^{19937}\!-\!1$ into $M$ sub-sequences of length
$$J=\myceil{\frac{P}{M}}\approx 2^{19937-\log_2 M}$$
with the last sub-sequence having length $J\!-\!1$. Table~\ref{tab:jvalues} shows the parameters $L$, $M$ and $J$ for all three V-family generators across the supported ISA widths.

\begin{table}[h!]
	\centering
	\small
    \renewcommand{\arraystretch}{1.2}
	\begin{tabular}{|c|c|c|c|c|c|c|c|c|c|}
		\hline
		& \multicolumn{3}{c|}{V-MT19937} & \multicolumn{3}{c|}{V-MT19937-64} & \multicolumn{3}{c|}{V-SFMT19937} \\
		\cline{2-10}
		ISA & $L$ & $M$ & $J$ & $L$ & $M$ & $J$ & $L$ & $M$ & $J$ \\
		\hline
		SSE2 or NEON & 128 & 4 & $2^{19935}$ & 128 & 2 & $2^{19936}$ & \multicolumn{3}{c|}{n.a.$^{*}$} \\
		\hline
		AVX2 or SVE256 & 256 & 8 & $2^{19934}$ & 256 & 4 & $2^{19935}$ & 256 & 2 & $2^{19936}$ \\
		\hline
		AVX-512 & 512 & 16 & $2^{19933}$ & 512 & 8 & $2^{19934}$ & 512 & 4 & $2^{19935}$ \\
		\hline
	\end{tabular}
	\caption{\label{tab:jvalues} V-family parameters $L$ (bits), $M$, and $J$ for the three generator types across ISA widths. $^{*}$V-SFMT19937 requires $L>128$ since SFMT state elements are 128-bit wide, making $M=1$ degenerate.}
\end{table}

A V-family generator producing the sequence $\boldsymbol{S}$ in \eqref{eq:combseq} is constructed by duplicating $M$ times the base MT generator with initial state $X_0$, yielding an array of $M$ generators $G_t$ ($t=0,\dots, M-1$) with identical state $X_0$. The state vector of each $G_t$ is then advanced by $t\, J$ steps so that it produces sub-sequence $\boldsymbol{S_t}$. Finally, the $M$ generators are combined into a single V-family generator that polls each $G_t$ in round-robin order.

Let's add a second index $t$ to the state vector, so that $X_{k,t}$ is the state $k$-th for the $t$-th generator, i.e. $X_{0,t}=X_{tJ}$, the multi-generator described below has initial state
$$
\begin{pmatrix}
    X_{0,0} \\ X_{0,1} \\ \vdots \\ X_{0,M-1}
\end{pmatrix}
=
\begin{pmatrix}
    x_{0,0} & x_{1,0} & \cdots & x_{n-1,0} \\
    x_{0,1} & x_{1,1} & \cdots & x_{n-1,1} \\
    \cdots \\
    x_{0,M-1} & x_{1,M-1} & \cdots & x_{n-1,M-1} \\
\end{pmatrix}
$$
By interleaving the words of the state vectors of the $M$ generators, we obtain for the multi-state generator a combined state vector $\hat{X}_0$
$$
    \hat{X}_0 = [(x_{0,0}, x_{0,1}, \dots x_{0,M-1}), (x_{1,0}, x_{1,1}, \dots x_{1,M-1}), \dots (x_{n-1,0}, x_{n-1,1}, \dots x_{n-1,M-1})]
$$
This allows to transform each of the $w$-bit operations in \eqref{eq:step} into an $L$-bit SIMD operation that advances all $M$ state vectors by one step simultaneously, reducing computation cost by a factor $M$. Using MT19937-32 as an illustration:
\begin{align}
\label{eq:msmt19937}
\begin{pmatrix}x_{k,0} \\ x_{k,1} \\ \vdots \\ x_{k,M-1} \end{pmatrix}
&= \begin{pmatrix}x_{k+m, 0} \\ x_{k+m,1} \\ \vdots \\ x_{k+m,M-1} \end{pmatrix} ~\XOR~ \left(\begin{pmatrix}x_{k,0} \\ x_{k,1} \\ \vdots \\ x_{k,M-1} \end{pmatrix}~\AND~h ~\OR~ \begin{pmatrix}x_{k+1,0} \\ x_{k+1,1} \\ \vdots \\ x_{k+1,M-1}  \end{pmatrix}~\AND~l\right) \star A
\end{align}
The analogous construction applies to MT19937-64 (with 64-bit lanes) and SFMT19937 (with 128-bit lanes), in each case replacing the scalar operation of \eqref{eq:step} or \eqref{eq:sfmt_step} with an $L$-bit SIMD operation across $M$ independent states. Because all $M$ states are advanced together in lock-step, the combined state $\hat{X}$ is always memory-contiguous and aligned, completely eliminating the alignment obstacle of Section~\ref{sec:vectorization}. The residual-iteration problem is also resolved: for a single state $n$ is not a multiple of $L/w$, but since each SIMD operation now advances all $M$ states by one step, the total work is exactly $n$ SIMD operations with no scalar remainder.

\subsection{Initialization via jump-ahead}
To construct the V-family generator described above, we need to advance the state vector $X_0$ by $J$ steps for each of the $M$ instances. In theory, we could apply recurrence \eqref{eq:step} $J$ times. However, in practice, since $J$ is a huge number, this operation would be astronomically expensive.
To overcome this, techniques known as \textit{jump ahead} are used to compute the state $X_{tJ}$ directly from $X_0$ without calculating all the intermediate states.

\subsubsection{\textit{Jump ahead} via binary matrix multiplications modulo 2}
\label{sec:jumpahead}
The simplest \textit{jump ahead} methodology for the MT19937 is based on binary matrix multiplications modulo 2, as proposed by Knuth \cite{knuth}. Let $Y$ be a subset of the state vector $X$, with the first $r$ bits (where $r = 31$) removed since they are not used in recurrence \eqref{eq:step}.
The effective state vector $Y$ is a binary matrix with dimensions $1\times (nw-r)$. Advancing the effective state vector $Y_k$ to $Y_{k+1}$ using \eqref{eq:step} is equivalent to the binary vector-matrix multiplication modulo-2:
\begin{equation}
Y_{k+1} = Y_k \star F
\end{equation}
where $F$ is a square binary matrix of size $wn-r$
\begin{equation}
\label{eq:transmat}
    F = \begin{pmatrix}
        0 & I_w & 0 & \cdots & &  &  &  & 0 \\
        \vdots & 0 & I_w & 0 & \cdots & & & & 0\\
        0 & \vdots & 0 & \ddots & &&& & \vdots \\
        I_w & 0 & \vdots & & \ddots & && & \vdots \\
        0 & \vdots &  & & & \ddots & & & \vdots \\
        \vdots &  &  &  & &&& I_w & 0 \\
        0 &  &  & &  & &&&I_{w - r} \\
        A & 0 & \cdots & &  &  &&& 0
\end{pmatrix}
\begin{matrix}
\\ \\ \\ \leftarrow m\text{-th row} \\ \\ \\ \\
\end{matrix}
\end{equation}
State $X_{tJ}$ can be computed as:
\begin{equation}
\label{eq:jumpahead}
    Y_{tJ} = Y_{tJ-1} \star F = Y_{tJ-2}  \star F \star F = \ldots = Y_{(t-1)J}  \star F^{J}, \quad j=1,\dots,M-1
\end{equation}
Let $B=F^J$, since $J$ is a power of 2 ($J=2^{q}$, where $q$ is the exponent from Table~\ref{tab:jvalues}).
The jump matrix $B$ is computed offline via repeated squaring $F^{2^q} = F^{2^{q-1}} \star F^{2^{q-1}}$ and stored for later use.

While the specific transition matrix structure $F$ in \eqref{eq:transmat} is representative of MT-based generators, the methodology of binary matrix exponentiation for jump-ahead is general and applies to any $F_2$-linear PRNG. A distinct jump matrix $B$ must be computed and stored for each of MT19937-32, MT19937-64, and SFMT19937, since their recurrences, word sizes, and state dimensions differ.

The overhead associated with initializing the generator state vectors as in \eqref{eq:jumpahead}, which requires $M\!-\!1$ vector-matrix multiplications, is negligible in the context of generating a large stream of PRNs.

\subsubsection{Alternative \textit{jump ahead} methods}
It is worth noting that there exist more efficient \textit{jump-ahead} algorithms that do not require storing the jump matrix and are faster than performing a vector-matrix multiplication as in \eqref{eq:jumpahead}. Examples of such algorithms can be found in \cite{jump1} and \cite{jump2}. These algorithms have not been discussed here, as the one adopted in section \ref{sec:jumpahead} is simpler to implement and is already very fast.
Furthermore, the main contribution of this paper is the vectorization of the algorithm to advance the state vector. If a more efficient \textit{jump-ahead} method were to be adopted, it would further improve initialization time, but it would not invalidate any of the arguments presented in this paper regarding the algorithm's throughput.

\section{Test results}

\subsection{Statistical tests}
As discussed in Section~\ref{sec:simdgen}, the V-family generators produce sequences of independent and identically distributed PRNs, since they combine sub-sequences of an underlying MT generator whose statistical properties are preserved by the interleaving. All three generators in the V-family, across all vectorization factors $M \in \{1, 4, 8, 16\}$, were subjected to the \textit{TestU01} test suite proposed by L'Ecuyer and Simard in \cite{testu01}. We observe that almost all tests pass, with results consistent with those of the corresponding original reference implementations. In particular, failures in the \textit{LinearComp} test are observed for all MT-based generators (including the original ones). These results demonstrate that the inter-state vectorization strategy preserves the statistical properties of the underlying base generators.

\subsection{Intra-State Vectorization: the X-Family}
\label{sec:xfamily}

The benchmarks presented in the next section include, in addition to the original reference implementations, the C++ STL and Intel MKL, a set of intra-state SIMD implementations implemented by the author: X-MT19937, X-MT19937-64, and X-SFMT19937. These apply SIMD instructions to vectorize the internal recurrence loop of a single generator instance, processing multiple consecutive loop iterations simultaneously within one state vector. Unlike the V-family, this approach preserves the exact sequential structure of the original algorithms: the output sequence is numerically identical to that of the corresponding reference implementation, making these generators direct drop-in replacements for existing code that relies on reproducible MT19937 output.

However, the intra-state approach remains subject to the structural obstacles identified in Section~\ref{sec:vectorization}: the loop iteration count is generally not a multiple of the SIMD width, requiring scalar fallback iterations for the remainder; and consecutive groups of state words are not memory-aligned, necessitating byte-shift or blend intrinsics to assemble the required operands. These limitations place a ceiling on the throughput of intra-state implementations that cannot be overcome without changing the algorithm's mathematical structure. The X-family is therefore included in the benchmarks to isolate the performance contribution of the V-family's novel inter-state parallelization strategy.

\subsection{Performance results}
\label{sec:tests}

We evaluated the performance of the full generator suite across five hardware targets to assess scaling over different SIMD registers length. Table~\ref{table:hardware} summarizes the hardware targets used in the benchmarks.

\begin{table}[H]
	\centering
	\small
	\begin{tabular}{|l|l|c|c|}
		\hline
		\textbf{Target ISA} & \textbf{CPU Model} & \textbf{Clock} & \textbf{L1d cache} \\
		\hline
		x86-64 SSE4.2  & Intel Celeron J4125 & 2.0 GHz & 24 KB \\
		x86-64 AVX2    & Intel Xeon E5-2686 v4 & 2.3 GHz & 32 KB \\
		x86-64 AVX-512 & Intel Xeon Platinum 8375C & 2.9 GHz & 48 KB \\
		ARM NEON       & Neoverse-N1 & 2.5 GHz & 64 KB \\
		ARM SVE256     & Neoverse-V1 & 2.6 GHz & 64 KB \\
		\hline
	\end{tabular}
	\caption{\label{table:hardware} Hardware targets.}
\end{table}

The benchmarks compare several algorithmic families:
\begin{itemize}
	\item the \textit{MT Family} (Original MT19937, STL MT19937, Intel MKL-MT, V-MT, and X-MT);
	\item the \textit{SFMT Family} (Original SFMT, Intel MKL-SFMT, V-SFMT, and X-SFMT);
	\item the \textit{MT-64 Family} (Original MT19937-64, STL MT19937-64, V-MT64, and X-MT64).
\end{itemize}	
We generate approximately $1.6 \times 10^{10}$ random numbers per hardware target (averaging 30 runs of $8.19 \times 10^6$ samples for each individual test) to ensure statistical stability.
Throughput is measured in Million samples per second (M/s). For 32-bit generators (MT, SFMT), samples are 32-bit words; for MT-64, samples are 64-bit words. Vectorial query mode performance is measured using a block size of 10,240 samples, which is a multiple of the generator state size.

To ensure a fair comparison in scalar mode, all scalar benchmarks use a \textit{volatile} variable trap within the query loop. This prevents the compiler from using Dead-Code Elimination (DCE) or auto-vectorization to skip or parallelize the PRNG operations, ensuring that the measured throughput reflects true sequential execution.

Speedup factors cited in the discussion below are computed relative to the corresponding original scalar implementation (ORIG-MT19937, ORIG-SFMT19937, or ORIG-MT19937-64) running on the same hardware.
The source code for the generators and the test framework is written in C++20 and is available on github\footnote{https://github.com/fabiocannizzo/VMT19937}. Tests were compiled with the GCC 16 compiler.

Performance test results are illustrated in Tables \ref{table:mt32}, \ref{table:sfmt}, and \ref{table:mt64}.

\begin{table}[H]
\centering
\small
\begin{tabular}{|l|l|c|c|c|c|c|}
\hline
\textbf{Query} & \textbf{Generator} & \textbf{SSE4.2} & \textbf{AVX2} & \textbf{AVX-512} & \textbf{NEON} & \textbf{SVE256} \\
\hline
\multirow{5}{*}{Scalar}
 & ORIG-MT19937 & 160.5 & 258.7 & 418.0 & 226.7 & 362.3 \\
 & STL-MT19937  & 134.2 & 305.7 & 598.6 & 276.8 & 569.8 \\
 & MKL-MT19937\textsuperscript{*} & 23.4 & 40.7 & 52.9 & n.a. & n.a. \\
 & \textbf{X-MT19937} & 317.2 & 417.1 & 929.5 & 420.9 & 671.7 \\
 & \textbf{V-MT19937} & \textbf{341.3} & \textbf{450.7} & \textbf{1354.0} & \textbf{431.4} & \textbf{687.7} \\
\hline
\multirow{5}{*}{Vectorial}
 & ORIG-MT19937\textsuperscript{†} & 160.5 & 258.7 & 418.0 & 226.7 & 362.3 \\
 & STL-MT19937\textsuperscript{†}  & 134.2 & 305.7 & 598.6 & 276.8 & 569.8 \\
 & MKL-MT19937 & 710.7 & 1667.6 & \textbf{6578.9} & n.a. & n.a. \\
 & \textbf{X-MT19937} & 710.9 & 1968.8 & 6504.6 & 582.7 & 1617.8 \\
 & \textbf{V-MT19937} & \textbf{857.9} & \textbf{2391.8} & 5841.4 & \textbf{634.0} & \textbf{1869.1} \\
\hline
\end{tabular}
\caption{\label{table:mt32} MT19937 Family Performance (M/s).
	(*) MKL scalar tested in vectorial mode with block size 1.
	(†) No native vectorial support for ORIG and STL; results copied from the scalar case.
	}
\end{table}

\begin{table}[H]
\centering
\small
\begin{tabular}{|l|l|c|c|c|c|c|}
\hline
\textbf{Query} & \textbf{Generator} & \textbf{SSE4.2} & \textbf{AVX2} & \textbf{AVX-512} & \textbf{NEON} & \textbf{SVE256} \\
\hline
\multirow{4}{*}{Scalar}
 & ORIG-SFMT19937 & 439.8 & 791.5 & 1179.9 & 523.5 & 751.7 \\
 & MKL-SFMT19937\textsuperscript{*} & 20.6 & 38.0 & 62.1 & n.a. & n.a. \\
 & \textbf{X-SFMT19937} & \textbf{471.6} & 847.9 & 1193.0 & \textbf{555.2} & 671.8 \\
 & \textbf{V-SFMT19937} & n.a. & \textbf{943.9} & \textbf{1507.6} & n.a. & \textbf{898.3} \\
\hline
\multirow{4}{*}{Vectorial}
 & ORIG-SFMT19937 & 1443.6 & 2503.5 & 4445.2 & \textbf{1380.9} & 1692.1 \\
 & MKL-SFMT19937  & 1654.4 & 3360.3 & 2761.7 & n.a. & n.a. \\
 & \textbf{X-SFMT19937} & \textbf{1699.6} & 2308.4 & 3611.8 & 1234.4 & 1327.4 \\
 & \textbf{V-SFMT19937} & n.a. & \textbf{4514.8} & \textbf{10942.6} & n.a. & \textbf{2960.4} \\
\hline
\end{tabular}
\caption{\label{table:sfmt} SFMT19937 Family Performance (M/s).
	(*) MKL scalar tested in vectorial mode with block size 1.
    }
\end{table}

\begin{table}[H]
\centering
\small
\begin{tabular}{|l|l|c|c|c|c|c|}
\hline
\textbf{Query} & \textbf{Generator} & \textbf{SSE4.2} & \textbf{AVX2} & \textbf{AVX-512} & \textbf{NEON} & \textbf{SVE256} \\
\hline
\multirow{4}{*}{Scalar}
 & ORIG-MT19937-64 & 140.4 & 179.8 & 337.5 & 202.6 & 337.1 \\
 & STL-MT19937-64  & 163.8 & 316.7 & 590.1 & 285.0 & \textbf{572.6} \\
 & \textbf{X-MT19937-64} & 179.5 & 325.0 & 1095.4 & 284.1 & 555.8 \\
 & \textbf{V-MT19937-64} & \textbf{207.6} & \textbf{371.8} & \textbf{1134.6} & \textbf{333.8} & 513.2 \\
\hline
\multirow{4}{*}{Vectorial}
 & ORIG-MT19937-64\textsuperscript{†} & 140.4 & 179.8 & 337.5 & 202.6 & 337.1 \\
 & STL-MT19937-64\textsuperscript{†}  & 163.8 & 316.7 & 590.1 & 285.0 & 572.6 \\
 & \textbf{X-MT19937-64} & 251.6 & 827.3 & 2524.6 & 255.1 & \textbf{926.8} \\
 & \textbf{V-MT19937-64} & \textbf{378.3} & \textbf{1161.3} & \textbf{2864.5} & \textbf{289.5} & 900.6 \\
\hline
\end{tabular}
\caption{\label{table:mt64} MT19937-64 Family Performance (M/s).
	(†) No native vectorial support for ORIG and STL; results copied from the scalar case.
	}
\end{table}

\subsection{Discussion and Observations}
\label{sec:obs}

The X-family generators (X-MT19937, X-MT19937-64, X-SFMT19937) are numerically identical to the original reference implementations. They have been vectorized to the fullest extent possible subject to the structural limitations identified in Section~\ref{sec:vectorization}. As such, they serve as direct equivalent competitors to other intra-state SIMD implementations, such as those provided by Intel MKL, the auto-vectorization introduced by GCC for the STL MT implementation, or the reference SFMT implementation on 128-bit registers. Any performance differences observed between the X-family and these competing implementations are attributable to the different implementation techniques employed by the authors of the various libraries to leverage SIMD instructions.

\subsubsection{Scalar Performance and Optimized Buffering}
A notable observation from the scalar sections of Tables~\ref{table:mt32} and \ref{table:mt64} is that both V-* and X-* implementations for the 32-bit and 64-bit MT variants outperform the corresponding original generators by a factor of 1.5 to 3.5, while the SFMT variants (Table~\ref{table:sfmt}) show more modest speedups. This general improvement is due to a buffered SIMD tempering strategy technique implemented by the author: state recurrence and tempering are executed in bulk using SIMD instructions, and results are served from a small cache-aligned buffer, amortizing the cost of state regeneration across many queries.

In contrast, Intel MKL in scalar mode (tested by requesting blocks of size 1) exhibits substantially lower throughput than even the original reference implementations. This confirms that MKL is optimized exclusively for large-block throughput and incurs a significant per-call overhead that renders it unsuitable for fine-grained scalar access.

\subsubsection{Proportional Scaling with SIMD Register Width}
The vectorial sections of Tables~\ref{table:mt32}, \ref{table:sfmt}, and \ref{table:mt64} confirm that the throughput of V-family generators in vectorial query mode scales roughly proportionally to the SIMD register width.

For V-MT19937, vectorial throughput relative to the original grows from approximately $5\times$ at SSE4.2 (128-bit, $M=4$) to approximately $14\times$ at AVX-512 (512-bit, $M=16$), consistent with the 4-fold increase in the number of parallel states. Comparing across ISA widths within V-MT19937 itself, throughput roughly doubles with each doubling of register width, reflecting near-linear scaling in $M$.

For V-SFMT19937, throughput at AVX-512 is approximately $2.4\times$ that at AVX2, matching the expected factor of 2 increase in $M$ ($2\to4$), while V-SFMT19937 itself delivers up to $2.5\times$ the throughput (at AVX-512) of the original SFMT at the same ISA in vectorial mode.

For V-MT19937-64, throughput grows by approximately $7.6\times$ from SSE4.2 to AVX-512, consistent with the increase in $M$ from 2 to 8. Across all three variants, the observed scaling validates the core design goal of the V-family: throughput grows proportionally to the SIMD register width.

\subsubsection{Cache Pressure and Memory Footprint}
As the SIMD width increases, so does the combined state storage of the V-family. V-MT19937 with $M=16$ parallel states requires $16 \times 624 \times 4 = 39{,}936$ bytes $\approx 40$~KB of state storage, which, combined with the output storage, saturates or exceeds the L1 data cache on architectures with 32--48~KB caches. This leads to more frequent L2 accesses and explains why the vectorial-mode speedup for V-MT19937 at AVX-512 falls somewhat short of the ideal $16\times$ predicted by the state count alone.

The V-MT19937-64 ($M=8$, state $\approx 20$~KB) and V-SFMT19937 ($M=4$, state $\approx 10$~KB) generators have smaller footprints and are therefore less susceptible to cache pressure, which is reflected in their more consistent scaling behavior.

The X-family, by contrast, maintains a single state vector of $\approx 2.5$~KB (for both MT19937-32 and MT19937-64), which fits comfortably in L1 cache. On AVX-512 in vectorial query mode, X-MT19937 actually outperforms V-MT19937 (Table~\ref{table:mt32}) for this reason: the reduced memory pressure and superior cache locality of the smaller single-state vector more than compensate for the structural vectorization barriers.

This highlights a clear architectural trade-off: for generators with very large states, the V-family's perfect vectorization can be bottlenecked by L1 cache capacity at the widest SIMD widths. This is a limitation that may disappear in the future with the continuous hardware evolution trend and increase in CPU L1 cache size.

\section{Conclusion}
In this paper, we introduced the \textbf{V-family}, a new algorithmic framework for the inter-state SIMD vectorization of Mersenne Twister generators. The core idea --- combining $M$ independent MT instances de-phased via jump-ahead transformations and polled in round-robin order --- achieves perfect inter-state vectorization, fully eliminating the alignment and residual-iteration obstacles that limit intra-state approaches. Applied to three base generators (MT19937-32, MT19937-64, and SFMT19937), the resulting generators V-MT19937, V-MT19937-64, and V-SFMT19937 demonstrate throughput that scales roughly proportionally to the SIMD register width, yielding speedups of an order of magnitude over the corresponding original implementations in vectorial query mode.

This performance comes at the cost of a larger state size: for generators with large state vectors such as MT19937-32, V-family state storage can reach 40~KB at 512-bit width. In some circumstances, this increased memory footprint can introduce cache pressure and moderate the scaling advantage at the widest register sizes.

The methodology introduced here is not specific to the Mersenne Twister and generalizes naturally to any linear recurrence PRNG, offering a principled approach for exploiting modern SIMD hardware without sacrificing period, statistical guarantees, or ease of implementation.

The methodology extends naturally to devices with massive vectorization capabilities, such as GPGPUs, which might be the topic of further studies.

%------------------------------BIBLIOGRAPHY-----------------

\end{document}